\documentclass[journal,12pt,onecolumn,letterpaper]{IEEEtran}
\usepackage{arxiv}
\usepackage{geometry}
\usepackage{times}
\usepackage{cite}
\usepackage{url}
\usepackage{graphicx}
\graphicspath{{Images_SimplyMime/}}
\usepackage{lscape}
\usepackage{subfigure}
\usepackage{rotating}
\usepackage{rotfloat}
\usepackage{xcolor}
\usepackage{amsmath}
\usepackage{amssymb}
\usepackage[linesnumbered,ruled,vlined]{algorithm2e}
\usepackage{pseudocode}
\usepackage{array}
\usepackage[english]{babel}
\usepackage{gensymb}
\usepackage{textcomp}
\usepackage{placeins}
\usepackage{balance}
\usepackage{booktabs}

\title{Colaboot: A Cloud-based Diskless PC Booting Mechanism}

\author{%
  
    Aditya Mitra \\
    Centre of Excellence, Artificial Intelligence \& Robotics (AIR),\\
    School of Computer Science and Engineering\\
    VIT-AP University, India \\
    \texttt{adityaarghya0@gmail.com}
\And

Anisha Ghosh\\
    Centre of Excellence, Artificial Intelligence \& Robotics (AIR),\\
    School of Computer Science and Engineering\\
    VIT-AP University, India \\
    \texttt{ghoshanisha2002@gmail.con}
\And
  Sibi Chakkaravarthy Sethuraman\\
    Centre of Excellence, Artificial Intelligence \& Robotics (AIR),\\
    School of Computer Science and Engineering\\
    VIT-AP University, India \\
    \texttt{sb.sibi@gmail.com} \\
\And
    Devi Priya V S \\
    Department of Computer Science and Engineering\\
    Siddhartha Academy of Higher Education, India \\	
   \texttt{vsdevipriya@gmail.com}
}

\begin{document}



\maketitle
\begin{abstract}
  Recent increases in endpoint-based security events and threats compelled enterprise operations to switch to virtual desktop infrastructure and web-based applications. In addition to reducing potential hazards, this has guaranteed a consistent desktop environment for every user. On the other hand, the attack surface is greatly increased because all endpoints are connected to the company network, which could harbor malware and other advanced persistent threats.  This results in a considerable loss of system resources on each individual endpoint.  Hence our work proposes a standard called Colaboot that enables machines throughout a company to boot from a single operating system in order to address these problems and guarantee a consistent operating system environment that could be easily updated to the most recent security patches across all work stations.
\end{abstract}




\enlargethispage{10pt}

\section{Introduction}

A study by the Ponemon Institute found that 68\% of businesses had experienced one or more endpoint attacks that successfully compromised IT infrastructure or data \cite{news_2024}. According to the same study, 68\% of IT professionals contend that endpoint incidents have gone up over the previous year. One of the most common ways that hackers target endpoints is via malware, which can be installed on the victim's device in several ways \cite{fabian2007endpoint}.
With desktop virtualization, employees can connect remotely to on-premises or cloud-based infrastructure that houses virtualized business workstations, doing away with the requirement for them to physically possess company machines \cite{lee2017understanding}. By doing this, a company can let workers work remotely without compromising control over its systems and data. Data is usually stored on servers by Virtual Desktop Interface (VDI) systems instead of local endpoints. As sensitive data is kept on the server, there is less chance of data loss or theft if an endpoint device is compromised. Businesses frequently use  VDI to guarantee a uniform desktop experience across all terminals. It makes it possible to display the desktop remotely via terminal connections to virtual machines (VMs) running on enterprise servers. As a result, it is common to see that the terminal's native resources, including the storage, are primarily utilized to boot the operating system. Terminals with fewer physical resources are known as thin clients \cite{parichha2010remote} and are designed to reduce resource wastage. However, it poses the potential of becoming a point of entry for malware or other apps onto the client's computer and the company network. Supplying resources to the virtual desktop without over provisioning while cutting down on resource utilization and reaction time is the primary resource allocation goal in the VDI \cite{hwang2012adaptive}.

In contemporary computing systems,disk-based booting faces several challenges \cite{kim2021rocky}. 
When compared to solid-state disks (SSDs) or network-based techniques, disk-based booting —especially from conventional hard disk drives (HDD) -is slower. Physical wear and tear is a common occurrence for HDDs as they have movable components. This can cause failure and possibly data loss, making them less dependable for frequent boot operations. The use of disk-based booting could lead to security risks, particularly if the disk is not encrypted. Vulnerabilities might arise if unauthorized individuals gain physical access to the disk and extract confidential information or change the boot procedure. Disk-based booting could prove ineffective and challenging to administer in large-scale enterprise environments.  Without the need for individual disk maintenance, network booting (such as Preboot Execution Environment (PXE)) enables centralized control and distribution of OS systems across numerous workstations. The energy consumption of HDDs is higher than that of SSDs, which can be detrimental in settings where energy conservation is important, such as data centers or mobile devices. In portable devices, disk-based booting could reduce battery life due to its increased power consumption.  In the event that the boot disk fails, recovery may be difficult and time-consuming. Repetitive boot options, such as booting from a network, USB, or SSD, are frequently used in modern systems to improve system availability and speed of recovery.

An innovative approach for booting thin clients directly from cloud storage is described in the proposed framework. Any cloud-native infrastructure, such as an IPython notebook running on Google Colab, can be used to provision the operating systems. This makes it possible to use thin clients without any storage media. With read-only rights, the operating system is streamed straight from the cloud, making the entire infrastructure resistant to Advanced Persistent Attacks (APTs). Since malware can only remain in volatile memory (RAM) \cite{sihwail2018survey}, even if it infects a client, the machine will be completely cleaned up when it is powered down. Further, it facilitates very smooth system upgrades and recovery.

\begin{table*}[!htbp]
\caption{Comparative Analysis with State-of-the-Art Techniques}
\label{Comparison}
\begin{tabular}{|l|l|l|l|l|l|l|l|}
\hline
                                                                               & \multicolumn{1}{c|}{\textbf{Boot Source}}                                & \multicolumn{1}{c|}{\textbf{Dependancy}}                              & \multicolumn{1}{c|}{\textbf{\begin{tabular}[c]{@{}c@{}}Hardware \\ Requirement\end{tabular}}} & \multicolumn{1}{c|}{\textbf{Scalability}}                                          & \multicolumn{1}{c|}{\textbf{Security}}                                       & \multicolumn{1}{c|}{\textbf{Cost}}                                                              & \multicolumn{1}{c|}{\textbf{Backup}}                                             \\ \hline
\textbf{\begin{tabular}[c]{@{}l@{}}Traditional \\ Booting\end{tabular}}        & \begin{tabular}[c]{@{}l@{}}Internal HDD/\\ SDD\end{tabular}              & \begin{tabular}[c]{@{}l@{}}Requires \\Local\\ Storage\end{tabular}      & BIOS/ UEFI                                                                                    & \begin{tabular}[c]{@{}l@{}}Limited by\\ Physical \\ Hardware\end{tabular}          & \begin{tabular}[c]{@{}l@{}}Data is stored \\ locally\end{tabular}            & \begin{tabular}[c]{@{}l@{}}High Initial\\  Investment\end{tabular}                              & \begin{tabular}[c]{@{}l@{}}Manual \\ Setting \\ up\end{tabular}                  \\ \hline
\textbf{\begin{tabular}[c]{@{}l@{}}Network \\ Booting\\ (PXE) \end{tabular}}    & \begin{tabular}[c]{@{}l@{}}Network Server\\ via PXE\end{tabular}         & \begin{tabular}[c]{@{}l@{}}Network \& \\ PXE Server\end{tabular}      & \begin{tabular}[c]{@{}l@{}}PXE capable\\ network card\end{tabular}                            & \begin{tabular}[c]{@{}l@{}}Limited by\\ network and\\ server capacity\end{tabular} & \begin{tabular}[c]{@{}l@{}}Firewalls, IPS,\\ Anitiviruses\end{tabular} & \begin{tabular}[c]{@{}l@{}}Cost of network \\ infrastructure\end{tabular}                       & \begin{tabular}[c]{@{}l@{}}Configured \\ with network \\ redundancy\end{tabular} \\ \hline
\textbf{\begin{tabular}[c]{@{}l@{}}USB/ \\External\\ Booting\end{tabular}}       & USB/HDD/SDD                                                              & \begin{tabular}[c]{@{}l@{}}External\\ devices\end{tabular}            & \begin{tabular}[c]{@{}l@{}}USB port and \\ compatible \\ external devices\end{tabular}        & \begin{tabular}[c]{@{}l@{}}Limited to \\ external device \\ capacity\end{tabular}  & Encrypted data                                                               & \begin{tabular}[c]{@{}l@{}}Low to \\ Moderate\end{tabular}                                      & \begin{tabular}[c]{@{}l@{}}Manual or \\ External\end{tabular}                    \\ \hline
\textbf{\begin{tabular}[c]{@{}l@{}}Recovery \\ Partition \\ Boot\end{tabular}} & \begin{tabular}[c]{@{}l@{}}Partition on \\ Internal Storage\end{tabular} & \begin{tabular}[c]{@{}l@{}}Recovery \\ Partition\end{tabular}         & \begin{tabular}[c]{@{}l@{}}Local storage \\ with partition\end{tabular}                       & \begin{tabular}[c]{@{}l@{}}Limited by \\ partition size\end{tabular}               & \begin{tabular}[c]{@{}l@{}}Built-in \\Recovery\\ tools\end{tabular}            & \begin{tabular}[c]{@{}l@{}}Included with \\ system\end{tabular}                                 & \begin{tabular}[c]{@{}l@{}}Built-in\\ recovery\end{tabular}                      \\ \hline
\textbf{Colaboot}                                                              & \begin{tabular}[c]{@{}l@{}}Cloud Remote\\ Disk\end{tabular}              & \begin{tabular}[c]{@{}l@{}}Stable internet \\ connection\end{tabular} & \begin{tabular}[c]{@{}l@{}}Minimal local \\ hardware\end{tabular}                             & Highly Scalable                                                                    & \begin{tabular}[c]{@{}l@{}}Cloud-based \\ security \\measures\end{tabular}     & \begin{tabular}[c]{@{}l@{}}Subscription \\based, potential\\ savings on \\ hardware\end{tabular} & \begin{tabular}[c]{@{}l@{}}Built-in\\ cloud\\ backup\end{tabular}                  \\ \hline
\end{tabular}
\end{table*}

\subsection{Contribution}
 We make the following contributions:
\begin{itemize}
    \item Our research provides a framework that enables enterprise nodes to boot directly from cloud storage, reducing the risk of persistent threats and facilitating smooth OS upgrades and maintenance.
    \item 
    To the best of our knowledge, Colaboot is the first and pioneering product of its kind that makes it possible for on-premise computers to boot from cloud storage. 
    \item We could experimentally confirm that Colaboot can run on PCs with Legacy BIOS and Unified Extensible Firmware Interface (UEFI) without the need for any physically attached disks or devices. 
\end{itemize}

\begin{sidewaysfigure*}[!htbp]
\centering
\includegraphics[width=1\textwidth]{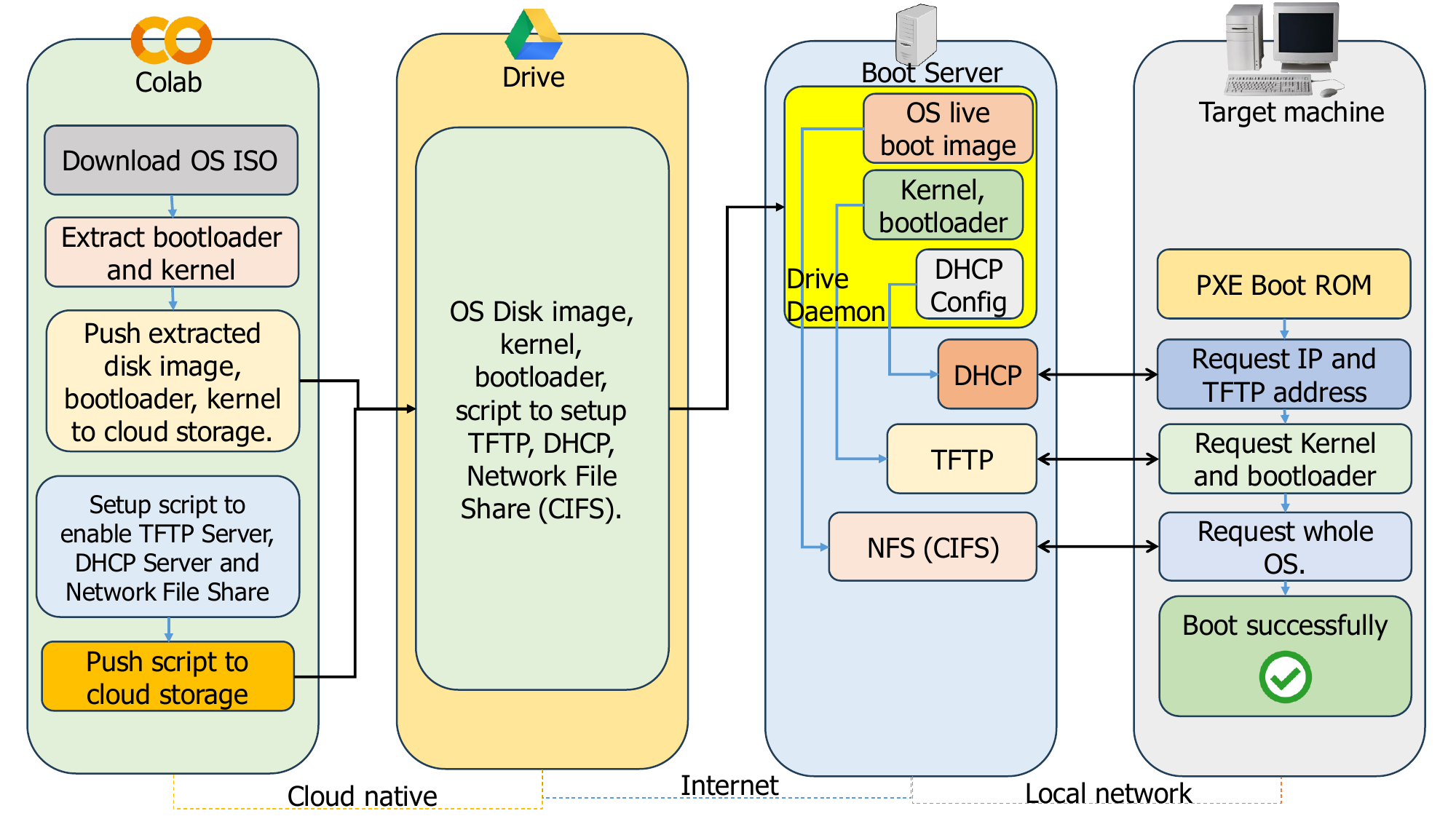}
\caption{Colaboot: The Proposed System}
\label{archi}
\end{sidewaysfigure*}

\section{Background and Related Works}
By employing Remote Desktop Protocol (RDP), the end user's device receives the virtual desktop that operates in a data center. For companies looking to go from traditional desktop to virtual desktops, resource allocation in the VDI has emerged as one of the major problems and concerns at the moment. Using varying workloads, Nakhai et.al \cite{nakhai2017performance} had assessed multiple operating systems as dedicated virtual desktops in the virtual desktop architecture to determine which OS had the lowest resource utilization and reaction time in terms of memory usage, CPU usage, and application response time.
A smooth migration depends on careful planning and testing, which is mentioned in the article \cite{stites2024legacy}, for enterprises that switch from outdated iSCSI to new technology.  It offers recommendations on how to satisfy contemporary data center requirements by utilizing the improved capabilities of the E810 controller, such as increased throughput, decreased latency, and sophisticated security choices. In order to reduce the requirement for local storage and desktop system management, the authors in \cite{cruz2010integration} addresses the architecture of PXE-based solutions and their benefits. However, not much has been researched about how PXE affects network efficiency and performance in really large-scale settings\cite{fuchs2023split}. Table \ref{Comparison} illustrates the analysis of existing booting techniques.

\section{System Model and Experimental Setup}

 The majority of contemporary computers are capable of installing an operating system directly onto the disk when the setup files are transferred over a local network owing to PXE capability \cite{fuchs2023split,guojie2020tpcm}. For the boot server to be accessed via the Boot Information Negotiation Layer (BINL), the network must have a Dynamic Host Control Protocol (DHCP) server. An open-source network boot server for learning environments was discussed by Aryotejo et.al. \cite{aryotejo2021open}. When a machine is configured to send a DHCPDISCOVER with PXEClient tag as a broadcast, the PXE boot procedure expands the DHCP protocol by adding the data required to remotely boot a computer. Following a request, the DHCP server returns a DHCPOFFER that includes the BINL Information, default gateway, IP address, subnet mask, and DNS server. Along with the filenames for the kernel and bootloader, the BINL Information also includes the IP address of the "Next Server." Subsequently, using Trivial File Transfer Protocol (TFTP), the machine tries to connect to the boot server, also known as the Next Server, and downloads the kernel and bootloader. Following kernel loading, it obtains the necessary data to load the operating system via Hypertext Transfer Protocol (HTTP), Common Internet File System (CIFS), or Network File Sharing (NFS). The device finishes the boot process by downloading the whole operating system into RAM. In enterprise settings, network boot is frequently employed to preserve uniformity in operating systems, security updates, and patches.

 In enterprise settings, network boot is frequently employed to preserve uniformity in operating systems, security updates, and patches.The boot server is typically hosted on an on-premise server, and the DHCP server must be present in the same subnet as the computer. A method for installing Linux over network booting was proposed in the work \cite{habibi2023installing}. A technique for diskless booting from local networks was presented by authors in \cite{suriya2022diskless}. Tiny Core Linux's reduced footprint, enhanced performance, and importance for network boot in control system situations are highlighted in the article \cite{washington2020migrating}, which also covers the advantages and procedure of switching control systems to Tiny Core Linux. The research by \cite{banik2022payload} demonstrated the creation of a pre-OS firmware block that leveraged network boot to identify system problems. To handle the Boot Server and operating systems properly, We propose a cloud-based technique that enables seamless updates, OS migrations, Zero Touch Provisioning  (ZTP), and other capabilities. By using an IPython notebook running Google Cloud disk, our solution provides Google Drive with the operating system data. It develops a deployment management script that is automatically deployed and executed on the Boot server.

\begin{figure}[h]
\centerline{\includegraphics[width=8cm,height=7cm]
{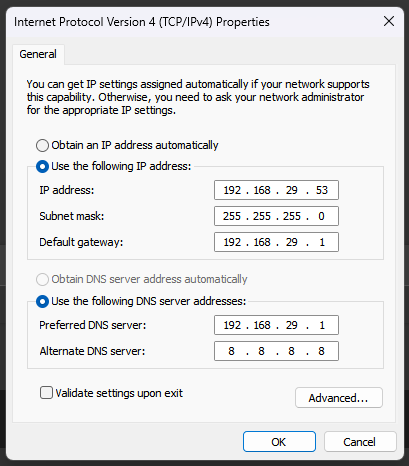}}
\caption{IP Configuration of Boot Server}
\label{Levels}
\end{figure}

Google Colab, a cloud-native environment, is employed for the suggested use case, along with an IPython notebook or Python script. It loads operating system disk images and stores boot files in a cloud storage service such as Google Drive. Google Drive directory is mounted as a TFTP and CIFS share by a Daemon that is operated by the boot server in the local datacenter. Pointing to the boot server as the "Next Server" is a DHCP server with BINL settings. Consequently, the Boot server's CIFS and TFTP shares would allow any computer connected to the network to boot from them. The framework and procedure of the suggested standard are depicted in Figure \ref{archi}. The boot server is configured with a static IP address since the setup configuration employs the boot server as the DHCP server. The network's other DHCP servers are maintained inactive. The IP setup of the boot server is displayed in Figure \ref{Levels}. 

\begin{figure}[htbp]
\centering
\includegraphics[width=1\textwidth]{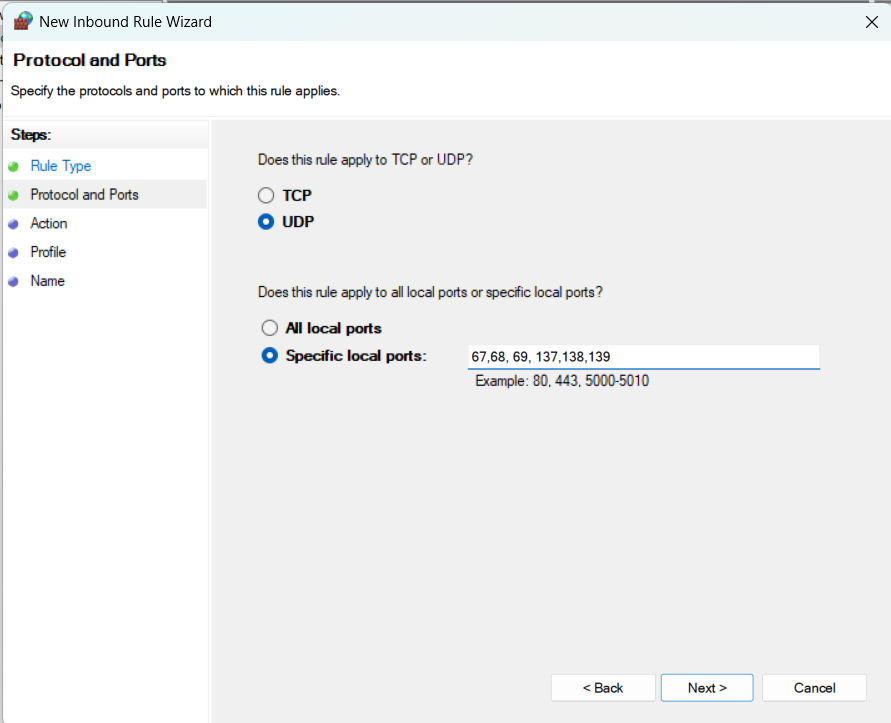}
\caption{Setup for Setting Firewall Rules}
\label{Setup  for Setting Firewall Rules}
\end{figure}

\begin{figure}[htbp]
\centering
\includegraphics[width=\columnwidth,height=6cm]{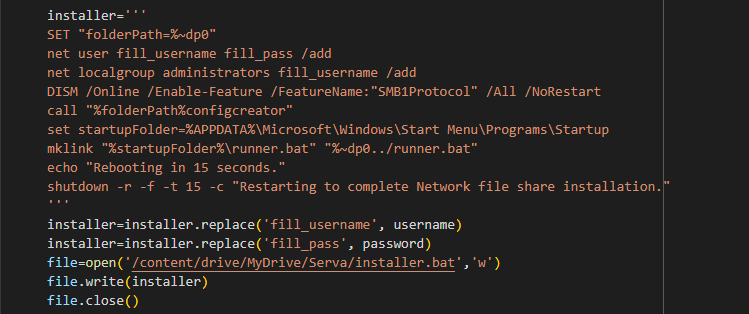}
\caption{Sample Window for the Generated Installer Script}
\label{Generating the installer script}
\end{figure}

During the process of configuring the static IP address, the boot server should retain internet connectivity. The Google Drive client is configured on the boot server to mount the Google Drive as a mount point. The Serva Community edition is then downloaded; it functions as a DHCP server, TFTP server, and BINL add-on server. To enable ports for TFTP (Port 69), DHCP (Ports 67, 68), and CIFS (Ports 137, 138,139, 445), we modified the firewall rules of the booting server. To accomplish these, navigate to Windows Security $\rightarrow$ Firewall $\rightarrow$ Advanced Settings $\rightarrow$ Set Inbound Rules $\rightarrow$  New Rule in the drop-down list. Figure \ref{Setup  for Setting Firewall Rules} illustrates the UDP inbound rules; correspondingly, TCP rules are configured. The notebook creates the Serva configuration files in the following stages, binding itself to the IP address of the bot server. Additionally, it generates a batch installation script whose purpose is to install SMB 1.0 support and CIFS on the Boot server, as well as create a user account for CIFS file sharing of the operating system. The installer script is generated, as shown in Figure \ref{Generating the installer script}.

Following the execution of this procedure, the downloaded Serva and created scripts are automatically synchronized with the boot server through the use of the Google Drive client. To set up the required services, the installer script can be executed on the boot server. The initial RAM disk (initrd) is created after the operating system has been installed and assets files, such as vmlinuz, pointing to the kernel image, are generated. Additionally, the configurations for the Boot server to connect to the network file share are added. Once the files are generated on Google Drive, it can be leveraged for booting.

\begin{sidewaysfigure*}[!htbp]
\centering
\includegraphics[width=1\textwidth]{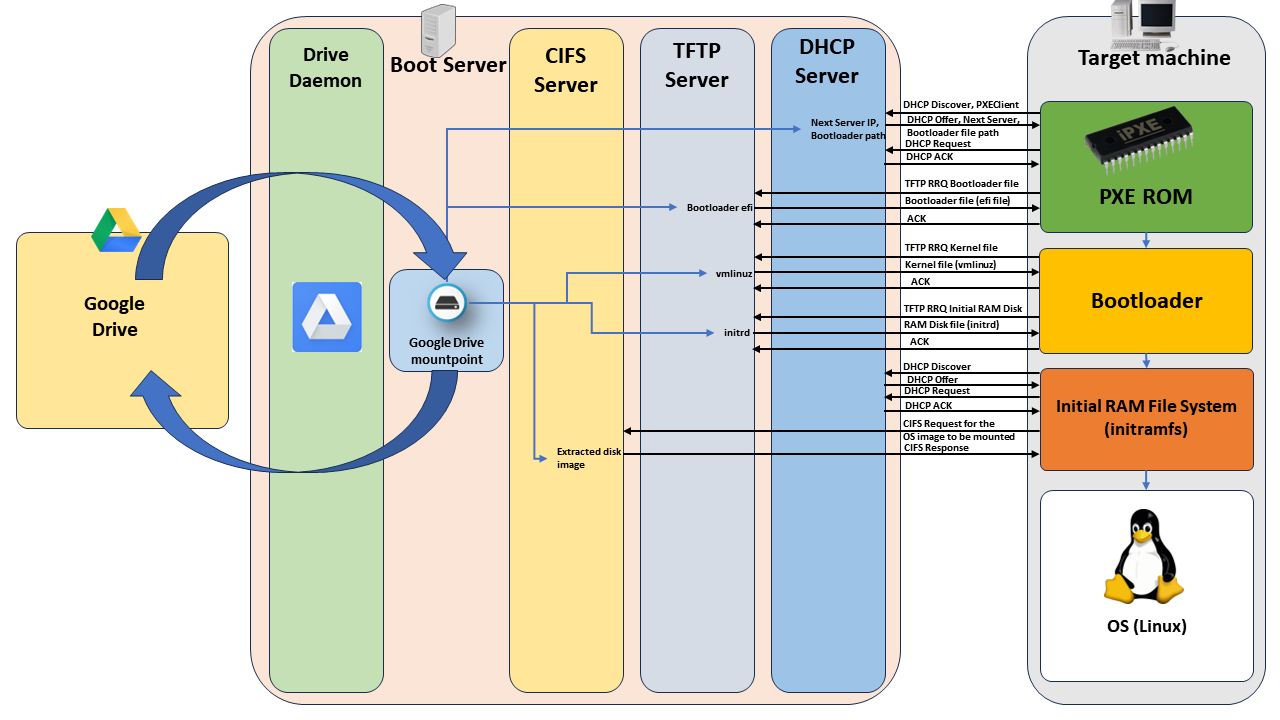}
\caption{Workflow of Booting Process}
\label{fig:workflow_of_booting_process}
\end{sidewaysfigure*}

\subsection{Booting Procedure}
The target machine uses the PXE Read Only Memory (ROM) during bootup. With the PXE Client add-on, it broadcasts a DHCP Discover. After receiving it, the DHCP server replies with the DHCP Offer with BINL add-on, which includes additional settings and the IP address of the following server. A DHCP ACK is sent back to the computer along with a DHCP Request.
Subsequently, the client computer requests the bootloader file from the TFTP server via a TFTP Read Request (RRQ). After receiving the response from the TFTP server, the machine loads the bootloader. Following that, the TFTP Server receives an RRQ from the bootloader requesting the compressed Kernel file. Additionally, the TFTP Server responds to an RRQ sent by the Bootloader for the Initial RAM Disk. The initial RAM file system is loaded from the Initial Ram disk as soon as it is received. The PC now has access to the CIFS server since it has a full network stack. It seeks to be booted live from the mount point where the full OS disk image from the CIFS server is mounted. When booting is finished, the desktop appears to the user. The Google Drive daemon streams all configurations and files needed for the bootloader, kernel, initial RAM disk, and disk image to the boot server. The booting process is depicted in Figure \ref{work Flow of Booting Process}. Consequently, the boot procedure enables the machine to be started using the operating system and configuration files stored on a cloud storage platform, such as Google Drive.

 \subsection{Experimental Setup Configuration}
  A Samsung Galaxy Book 3 running Windows 11 Pro is configured as the boot server, and it is connected via a USB-C to Ethernet NIC to a Dell Inspiron 3593 and VirtualBox as the target computers. The Google Drive for Desktop application is used to manage drives. The Serva Community Edition is used to administer TFTP and DHCP services. CAT 5E cables and a TP-Link 8-port Gigabit Ethernet switch provide wired connectivity, which guarantees dependable and effective network communication for the system. Results from the experimental arrangement were effective. However, the download of Ubuntu took more than five minutes to complete over CIFS, with an image size of about 4.1 GB. Additionally, a better boot server and network architecture could significantly increase speed as the USB C to Ethernet NIC utilized with the boot server posed a bottleneck.
\section{Use Case}
\subsection{Case Study 1:}In business settings where customers primarily utilize VDI or network-accessible services, Colaboot can be incredibly helpful. Since it uses the operating system in a read-only manner and removes the possibility of being impacted by Advanced Persistent Threats, it would also be advantageous from a security standpoint. The positive consequences of Colaboot are as follows:
\begin{itemize}
    \item When utilizing VDI, users frequently do not utilize the system's resources. Colaboot will thereby reduce the amount of resources needed and provide a consistent desktop environment for these terminals.
    \item Centralized updates and migrations: It is simple to modify the OS files of cloud disk notebook's from any authorized device, and the updated version takes effect the next time the enterprise's terminals power up.
    \item APT Mitigation: Read-only operating systems are put into RAM on the terminals. Because of this, any malware or APTs that are installed on the terminals will be automatically removed when the computer shuts off.
    \item Infrastructure as Code (IaC): In the IPython notebook, the operating system, the disk image, and other resources are referred to as Code, making it simple to adjust, add, or utilize different operating systems.
    \item  Instantaneous provisioning of controlled OS: Any controlled OS, or OS created for specific uses, can be uploaded to Google cloud disk, or any other notebook environment, and provisioned to all of the company's computers with ease.  The modifications will become effective with the first possible boot.
\end{itemize}
\subsection{Case Study 2:} Managing operating system installs, software upgrades, and data security presents difficulties for established universities with numerous computer labs that each house hundreds of PCs. The IT department of the university has to dedicate a substantial amount of time and money to the upkeep of these systems, frequently encountering challenges such as obsolete software, irregular system configurations, and data susceptibility resulting from local storage. Computers can boot from the cloud using Colaboot, eliminating local storage requirements. The IT staff could centrally manage and deploy a master image of the operating system and necessary software in the cloud, saving time and money compared to installing operating systems and apps on each PC separately. The positive con-
sequences of Colaboot are as follows:
\begin{itemize}
\item Users utilize a network login to authenticate themselves. A customized user profile loads after successful authentication, granting access to unique workspaces, preferences, and cloud-stored data.
    \item Manage OS and software centrally. The university may add more PCs without worrying about storage or software installation because of the system's scalability. As OS upgrades, patches, and software updates are deployed once in the cloud and appear on all PCs at once, centralized management lightens the burden on the IT staff.
    \item Every computer in the lab is set up to load the operating system image from the cloud and boot via the network (PXE boot).
    \item Data security is ensured since no data is kept locally; instead, all user files and data are saved straight to their cloud accounts. 
\end{itemize}
\subsection{Case Study 3:} Numerous offices, each housing hundreds of workstations, are operated by a multinational call center corporation in various countries. Agents answer client inquiries, manage support requests, and access customer databases using these workstations. Maintaining uniform workstation environments at all sites, controlling software upgrades, and stopping data leaks from local storage risks are issues the company must overcome. Data security could potentially improved and IT processes could be streamlined with Colaboot Mechanism. The organization could ensure a consistent operating environment throughout all locations and consolidate its IT management by switching to diskless workstations.
The company could host a common operating system image on a colaboot, together with all the required customer care software, such as ticketing systems, CRM tools, and communication apps. Every location has workstations configured to use the PXE boot mechanism to boot from the cloud. The workstation downloads the software and operating system image from the cloud when an agent login in, guaranteeing uniformity across all devices. The business may swiftly resume operations by swapping out the hardware and reestablishing its connection to the cloud in the event of a hardware malfunction or disaster at any location. The new workstations will be identical to the ones they replace because all configurations and data are saved in the cloud, reducing downtime.

 \section{Conclusion}
 Colaboot promises to improve resource usage among various terminals and give every user a uniform desktop experience in corporate environments. To establish Colaboot as a pioneering solution that allows on-premise machines to boot straight from the cloud, this article describes the deployment process and presents outcomes from experiments that demonstrate his efficacy. Colaboot does not require any physically linked disks or drives to operate on UEFI and Legacy BIOS-based systems. This will play a key role in enhancing enterprise terminals, reducing resource waste, and enhancing desktop environment consistency. Colaboot currently relies on a network and PXE server as key dependencies. The PXE server plays a crucial role in facilitating communication with the cloud, enabling the diskless booting process. However, this dependency introduces certain challenges, particularly in environments where network infrastructure might be unreliable or limited. Future iterations of Colaboot aim to overcome this limitation by exploring direct connections to cellular networks, potentially eliminating the need for a PXE server and further simplifying the deployment process. This advancement would enhance the flexibility and scalability of Colaboot, making it more adaptable to a wider range of operational environments. By creating an appropriate frontend and application to distribute the operating systems, this work can be further enhanced. Additionally, in the long run, it would be reasonable to start a system using several cloud storage options, such as AWS S3 buckets.

\section{Tutorial/CheatSheet and Demo}
The Colaboot cheatsheet can be accessed through the following GitHub repository \cite{colaboot2024}. A demo video of Colaboot is available at \cite{colaboot_demo}.

\section{Acknowledgments}

The authors would like to thank the editors and reviewers. The authors would like to thank Kota Reddy, Vice Chancellor, VIT-AP University, Jagadish Chandra Mudiganti, Registrar, VIT-AP University and Hari Seetha, Director, Centre of Excellence, Artificial Intelligence and Robotics (AIR)  for their support. Special thanks to the team members of the Artificial and Robotics (AIR) Center, VIT-AP University. %A similar version is archived \cite{Colaboot_arxiv}.

%\bibliographystyle{ACM-Reference-Format}

\bibliographystyle{elsarticle-num} 
\bibliography{Colaboot-arxiv}


\end{document}